\documentclass[graybox]{svmult}

\usepackage{mathptmx}
\usepackage{helvet}
\usepackage{courier}
\usepackage{type1cm}
\usepackage{makeidx}
\usepackage{graphicx}

\usepackage{multicol}
\usepackage[bottom]{footmisc}
\usepackage{bm}
\usepackage{amsmath}
\usepackage{stmaryrd}
\usepackage{hyperref}
\usepackage{amssymb}
\usepackage{multirow}
\usepackage{amsfonts}
\usepackage{subcaption}
\usepackage{upgreek}
\usepackage[outdir=./FiguresPDF/]{epstopdf}
\usepackage[hypcap]{caption}
\usepackage{xcolor}
\usepackage{soul}

\usepackage{etoolbox}

\providecommand{\ten}[1]{\mathrm{#1}}
\providecommand{\tenf}[1]{\mathbb{#1}}

\usepackage{floatrow}

\newfloatcommand{capbtabbox}{table}[][\FBwidth]

\makeindex

	\title*{Multiscale Modeling of Elasto-Plasticity in Heterogeneous Geomaterials Based on Continuum Micromechanics}
	\titlerunning{Multiscale Modeling of Elasto-Plasticity}
	
	\author{Mahdad Eghbalian, Mehdi Pouragha and Richard Wan}

	\institute{M. Eghbalian . R. Wan \at Dept. of Civil Engineering, University of Calgary, Calgary, Alberta, Canada. \\
		M. Pouragha \at Dept. of Civil and Environmental Engineering, Carleton University, Ottawa, Ontario, Canada.  \\
	M. Eghbalian \email{meghbali@ucalgary.ca}; M. Pouragha \email{mehdi.pouragha@carleton.ca};  R. Wan \email{wan@ucalgary.ca}}

\begin{document}

\maketitle

\abstract{In this paper, we investigate some micromechanical aspects of elasto-plasticity in heterogeneous geomaterials. The aim is to upscale the elasto-plastic behavior for a representative volume of the material which is indeed a very challenging task due to the irreversible deformations involved. Considering the plastic strains as eigen-strains allows us to employ the powerful tools offered by Continuum Micromechanics which are mainly developed for upscaling of eigen-stressed elastic media. The validity of such eigen-strain based formulation of multiscale elasto-plasticity is herein examined by comparing its predictions against Finite Element (FE) simulations.}

\section{Introduction}
Elastic properties, plastic deformations and failure strength of geomaterials are strongly influenced by their underlying heterogeneous microstructure, including shape and topology of individual material phases. Advances made in multiscale modeling techniques have allowed the development of macroscopic models for geomaterials that are predicated on the inherent microstructural features. Most notably, as an extension to classical linear homogenization techniques, Pichler and Hellmich \cite{Pichler2010} generalized the primitive Transformation Field Analysis (TFA) of Dvorak and Benveniste \cite{Dvorak1992a} developed for eigen-stressed elastic composite media, to arbitrary ellipsoidal shape and directional orientation of the material phases. For an elasto-plastic composite, on the other hand, complexities pertaining to the dissipative behavior of the material naturally arise; in particular, regarding the upscaling of plastic strains. The idea of considering the plastic strains as eigen-strains in homogenization was first proposed in the early 1990s (see \cite{Dvorak1992b}). Such an assumption appears to be reasonable given that the plastic strains are kinematically incompatible, thus qualify as free strains \cite{Morin2017}. As such, the homogenization of elasto-plasticity would be greatly simplified in that, the elasto-plastic material can be replaced with an equivalent eigen-stressed elastic media. Such an approach was adopted by Morin et al. \cite{Morin2017} within the generalized TFA framework of Pichler and Hellmich \cite{Pichler2010}; and later on by the authors \cite{Eghbalian2019} who extended the framework to model the poro-elasto-plastic behavior of clays. The current work examines the validity of the eigen-strain formulation of plastic strains by comparing the multiscale model predictions against FE results.

\section{Eigen-Strain Based Multiscale Elasto-Plasticity Framework}\label{sec:formulation}
Let us consider a heterogeneous Representative Elementary Volume (REV) of the material comprising elasto-plastic phases (denoted by the set $\mathcal{N}$). The phases generally have differing properties, shapes and directional orientations. The objective is to formulate the overall elasto-plastic constitutive relation of the REV by upscaling the local heterogeneous strain and stress fields within the REV.

\textit{Notations:} Mathematical double-struck capital letters denote fourth-order tensors, while second-order tensors and scalars are shown in normal type. Moreover, variables referring to the macroscopic scale are accented with an overline.

\subsection{Homogenization of Microscopic Elasto-Plasticity}
We assume all phases to show an elastic-perfectly plastic behavior and undergo small strains. A fundamental assumption is considering the plastic strains as eigen-strains. Under these conditions, the stress state inside the REV can be written as:
\begin{equation}\label{eq:REV-stress}
	\ten{\sigma}_{\upalpha}=\tenf{C}_{\upalpha} : \ten{\varepsilon}_{\upalpha}+\ten{\varsigma}_{\upalpha}\quad \forall\upalpha \in \mathcal{N}
\end{equation}
where $\ten{\sigma}$ and $\ten{\varepsilon}$ are stress and strain tensors, $\tenf{C}$ is the elasticity tensor, and $\ten{\varsigma}=-\tenf{C} : \ten{\varepsilon}^{\text{p}}$ is the eigen-stress tensor. The evolution of plastic strains ($\ten{\varepsilon}^{\text{p}}$) is assumed to follow the flow rule:
\begin{equation}\label{eq:flow-rule}
	\dot{\ten{\varepsilon}}^{\text{p}}=\dot{\lambda} \frac{\partial G}{\partial \ten{\sigma}}
\end{equation}
where the plastic potential $G$ is generally different from the yield function. The yield function $F$ and the plastic multiplier $\lambda$ satisfy the Melan-Kuhn-Tucker conditions. 

In the constitutive relation (Eq.~\eqref{eq:REV-stress}), the microscopic stress $\ten{\sigma}$ satisfies the momentum balance condition, while the strain $\ten{\varepsilon}$ is kinematically compatible. Moreover, the REV is subjected to uniform macroscopic strain $\overline{\ten{\varepsilon}}$ at its boundary (Hashin boundary condition). Such boundary condition guarantees that the volume average of strains inside the REV is equal to the prescribed macroscopic strain. Upon using the Levin's theorem for pre-stressed heterogeneous composite media, the macroscopic stress is obtained as:
\begin{equation}\label{eq:avg-stress}
	\overline{\ten{\sigma}}=\overline{\tenf{C}} : \overline{\ten{\varepsilon}}+\overline{\ten{\varsigma}}
\end{equation}
where the macroscopic elasticity $\overline{\tenf{C}}$ and eigen-stress $\overline{\ten{\varsigma}}$ are defined in terms of average properties of phases and a strain concentration tensor $\tenf{A}$ (see Section \ref{sec:return}) as:
\begin{subequations}
	\begin{alignat}{2}
		\overline{\tenf{C}}&=\sum_{\upalpha \in \mathcal{N}} f_{\upalpha}~ \tenf{C}_{\upalpha} : \tenf{A}_{\upalpha} \\
		\overline{\ten{\varsigma}}&=\sum_{\upalpha \in \mathcal{N}} f_{\upalpha}~ \tenf{A}^{\top}_{\upalpha} : \ten{\varsigma}_{\upalpha}=-\sum_{\upalpha \in \mathcal{N}} f_{\upalpha}~ \tenf{A}^{\top}_{\upalpha} : \tenf{C}_{\upalpha} : \ten{\varepsilon}_{\upalpha}^{\text{p}}
	\end{alignat}
\end{subequations}

In the above relations, $f$ denotes the volume fraction. Equation~\eqref{eq:avg-stress} can be re-stated in form of the classical constitutive relation for an elastic-perfectly plastic material:
\begin{equation}\label{eq:avg-stress-2}
	\overline{\ten{\sigma}}=\overline{\tenf{C}} : \left(\overline{\ten{\varepsilon}}-\overline{\ten{\varepsilon}}^{\text{p}}\right)
\end{equation}
with the following definition for the macroscopic ``plastic" strain $\overline{\ten{\varepsilon}}^{\text{p}}$:
\begin{equation}\label{eq:avg-pl-strain}
	\overline{\ten{\varepsilon}}^{\text{p}}={\overline{\tenf{C}}}^{-1} : \left(\sum_{\upalpha \in \mathcal{N}} f_{\upalpha}~ \tenf{A}^{\top}_{\upalpha} : \tenf{C}_{\upalpha} : \ten{\varepsilon}_{\upalpha}^{\text{p}}\right)
\end{equation}

\subsection{Multiscale Return-Mapping Algorithm}\label{sec:return}
In this section, we present the procedure for numerical integration of the multiscale elasto-plasticity equations over discrete time steps. Under applied loading, the state variables of the REV evolve, including the induced plastic strains at the microscopic scale. Tracking the evolution of microscopic plastic strains requires access to a localization relation for the strain inside the REV. Based on the generalized TFA formulation \cite{Pichler2010}, the phase averaged strains of the pre-stressed media can be related to the boundary strain and all the eigen-stresses occurring in all the other phases through employing the so-called concentration and influence tensors $\tenf{A}$ and $\tenf{B}$ as:
\begin{equation}\label{eq:strain-localization}
	\ten{\varepsilon}_{\upalpha}=\tenf{A}_{\upalpha} : \overline{\ten{\varepsilon}}-\sum_{\upbeta \in \mathcal{N}} \tenf{B}_{\upalpha \upbeta} : {\tenf{C}}^{-1}_{\upbeta} : \ten{\varsigma}_{\upbeta}=\tenf{A}_{\upalpha} : \overline{\ten{\varepsilon}}+\sum_{\upbeta \in \mathcal{N}} \tenf{B}_{\upalpha \upbeta} : \ten{\varepsilon}^{\text{p}}_{\upbeta}\quad \forall\upalpha \in \mathcal{N}
\end{equation}

Closed-form expressions for the concentration and influence tensors are presented in \cite{Eghbalian2019,Morin2017,Pichler2010} for different homogenization schemes. A fully strain-controlled loading scheme is considered here, while the extension to hybrid stress-strain loading can be found in \cite{Eghbalian2019}. Knowing the REV state at time step $n$ ($\ten{\varepsilon}^n$, $\ten{\varepsilon}^{\text{p},n}$, $\ten{\sigma}^{n}$, $\overline{\ten{\varepsilon}}^{n}$, $\overline{\ten{\varepsilon}}^{\text{p},n}$ and  $\overline{\ten{\sigma}}^{n}$) and given the increment in the macroscopic strain ($\dot{\overline{\ten{\varepsilon}}}^{n+1}$), the aim is to obtain the state variables at step $n+1$, i.e. $\ten{\varepsilon}^{n+1}$, $\ten{\varepsilon}^{\text{p},n+1}$, $\ten{\sigma}^{n+1}$, $\overline{\ten{\varepsilon}}^{n+1}$, $\overline{\ten{\varepsilon}}^{\text{p},n+1}$ and $\overline{\ten{\sigma}}^{n+1}$. For the macroscopic strain, we can write:
\begin{equation}\label{eq:macro-strain-update}
	\overline{\ten{\varepsilon}}^{n+1}=\overline{\ten{\varepsilon}}^{n}+\dot{\overline{\ten{\varepsilon}}}^{n+1}
\end{equation}

Next, in a trial attempt, we assume no additional microscopic plastic strains are induced in the REV during the current step. The trial microscopic strains and stresses at step $n+1$ are thus calculated for each phase $\upalpha$ using the localization relation (Eq.~\eqref{eq:strain-localization}) and constitutive relation (Eq.~\eqref{eq:REV-stress}) as:
\begin{subequations}\label{eq:trial-strain:trial-stress}
	\begin{alignat}{2}
		\varepsilon_{\upalpha}^{n+1,\text{tr}}&=\tenf{A}_{\upalpha} : \overline{\ten{\varepsilon}}^{n+1}+\sum_{\upbeta \in \mathcal{N}} \tenf{B}_{\upalpha \upbeta} : \ten{\varepsilon}^{\text{p},n}_{\upbeta}\quad \forall\upalpha \in \mathcal{N} \label{eq:trial-strain} \\
		\ten{\sigma}_{\upalpha}^{n+1,\text{tr}}&=\tenf{C}_{\upalpha} : \left(\ten{\varepsilon}_{\upalpha}^{n+1,\text{tr}} - \ten{\varepsilon}^{\text{p},n}_{\upalpha}\right)\quad \forall\upalpha \in \mathcal{N} \label{eq:trial-stress}
	\end{alignat}
\end{subequations}

Next, the assumption of purely elastic behavior (trial attempt) is checked for all phases via inserting the trial stresses into the yield criterion $F$:
\begin{equation}
	\mathcal{F}_{\upalpha}^{\text{tr}}=F\left(\ten{\sigma}_{\upalpha}^{n+1,\text{tr}}\right)\quad \forall\upalpha \in \mathcal{N}
\end{equation}

If $\mathcal{F}_{\upalpha}^{\text{tr}}<0 ~~\forall\upalpha \in \mathcal{N}$, the REV is elastic and the calculated trial strains and stresses are accepted. If for at least one phase $\mathcal{F}^{\text{tr}}>0$, the REV is plastic and return mapping should be performed. Referring to Eqs.~\eqref{eq:REV-stress}, \eqref{eq:strain-localization} and \eqref{eq:trial-strain:trial-stress}, the phase strains and stresses at step $n+1$ can generally be stated based on their trial values as:
\begin{subequations}\label{eq:update}
	\begin{alignat}{2}
		\ten{\varepsilon}_{\upalpha}^{n+1}&=\ten{\varepsilon}_{\upalpha}^{n+1,\text{tr}}+\sum_{\upbeta \in \mathcal{N}} \dot{\lambda}_{\upbeta}^{n+1} ~\tenf{B}_{\upalpha \upbeta} : \frac{\partial G}{\partial \ten{\sigma}_{\upbeta}}\quad \forall\upalpha \in \mathcal{N} \\
		\ten{\sigma}_{\upalpha}^{n+1}&=\ten{\sigma}_{\upalpha}^{n+1,\text{tr}}+\tenf{C}_{\upalpha} : \left(\sum_{\upbeta \in \mathcal{N}} \dot{\lambda}_{\upbeta}^{n+1} ~\tenf{B}_{\upalpha \upbeta} : \frac{\partial G}{\partial \ten{\sigma}_{\upbeta}}-\dot{\lambda}_{\upalpha}^{n+1}\frac{\partial G}{\partial \ten{\sigma}_{\upalpha}}\right)\quad \forall\upalpha \in \mathcal{N}
	\end{alignat}
\end{subequations}

We denote the set of elastic and plastic phases by $\mathcal{N}_\text{e}$ and $\mathcal{N}_\text{p}$, respectively. For the elastic phases we have $\dot{\lambda}_{\upalpha}^{n+1}=0 ~~\forall\upalpha\in\mathcal{N}_\text{e}$, while for the plastic phases $\dot{\lambda}^{n+1}$ is determined by satisfying the yield criterion:
\begin{equation}\label{eq:sys-eq}
	F\left(\ten{\sigma}_{\upalpha}^{n+1}\right)=0\quad \forall\upalpha \in \mathcal{N}_\text{p}
\end{equation}
resulting in a system of nonlinear equations that can be solved using the Newton iterative scheme. The calculated plastic multipliers are then used in Eq.~\eqref{eq:update} to update the microscopic stresses and strains, from which the macroscopic stresses and strains can be updated using Eqs.~\eqref{eq:avg-stress-2} and \eqref{eq:avg-pl-strain}. It should be pointed out that upon updating the microscopic stresses, it is possible that some of the phases that were initially deemed elastic also become plastic. It is also possible that the converged plastic multipliers become negative for some phases, which means those phases are no longer plastic. In both cases, the set of plasticized phases should be updated in an iterative process until convergence.

\section{Validation Against FE Results}
\begin{figure}[t]
	\begin{floatrow}
		
		\capbtabbox{%
			\caption{Material properties of the sample REV.}
			\centering
			\begin{tabular}{c | l c}
				& Parameter & Value \\ \hline
				\multirow{2}{*}{Matrix} & Young's modulus (MPa) & $100$ \\
				& Poisson's ratio & $0.25$ \\
				\hline
				\multirow{4}{*}{Inclusions} & Young's modulus (MPa) & $1000$ \\
				& Poisson's ratio & $0.25$ \\
				& $\phi$ (radian) & $0$ \\
				& $\ten{\sigma}_{0}$ (MPa) & $0.12$ \\
				\hline
			\end{tabular}
			\label{table:properties}
			\renewcommand{\arraystretch}{1}
		}{%
		}
		
		\ffigbox{%
			\begin{center}
				\includegraphics[width=0.45\linewidth]{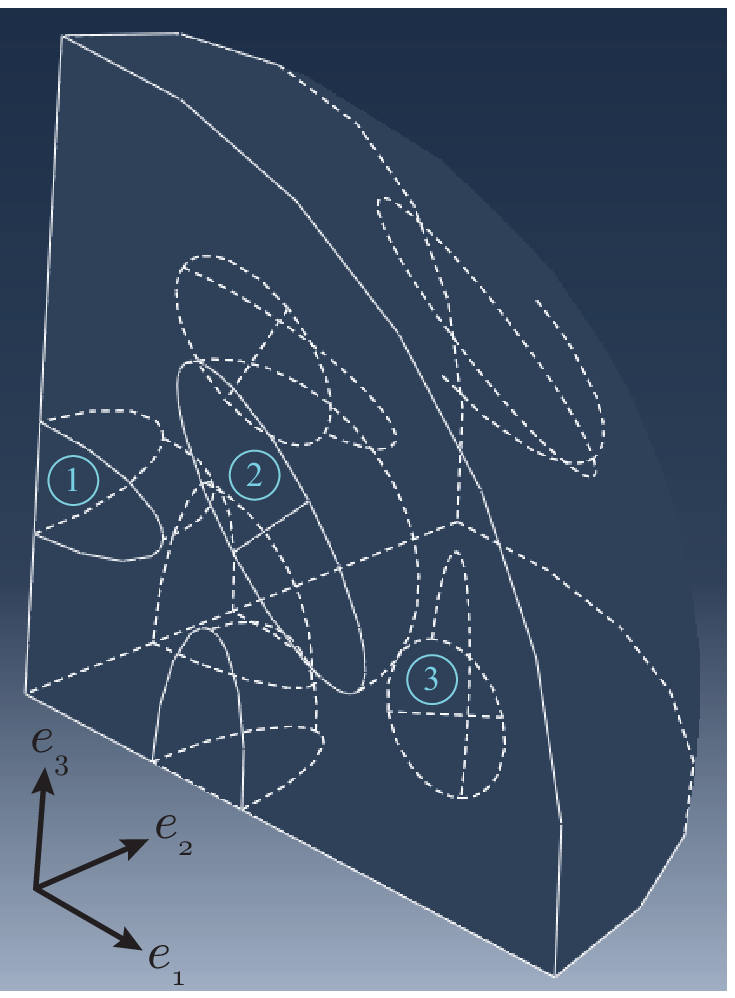}
				\caption{\label{fig:rev} $1/8$ of the sample REV used for FE simulations.}
			\end{center}%
		}{%
		}
	\end{floatrow}
\end{figure}

In this section, we validate the predictions of the multiscale formulation (section~\ref{sec:formulation}) against FE simulations where multiple phases are explicitly modeled. A 3D spherical REV of unit radius is considered which includes 26 oblate spheroidal inclusions with aspect ratio of $0.35$ and total volume fraction of $0.143$ distributed isotropically inside a matrix. The matrix is elastic, while the inclusions obey an associated Drucker-Prager elasto-plasticity:
\begin{equation}\label{eq:DP}
	F=G=\ten{\sigma}_{\text{eq}}+\ten{\sigma}_{\text{m}} \tan{\phi}-\ten{\sigma}_{0}
\end{equation}
where $\ten{\sigma}_{\text{m}}$ is the mean stress, $\ten{\sigma}_{\text{eq}}$ is the equivalent deviatoric stress, $\phi$ is the friction angle and $\ten{\sigma}_{0}$ is the failure stress under pure shear. The material properties of the REV are listed in Table~\ref{table:properties}. The REV is subjected to uniaxial compression loading by prescribing a total vertical strain of $-0.001$ and zero lateral stresses, followed by unloading to half the total vertical strain. For the eigen-stress based multiscale analysis, the Mori-Tanaka scheme is used for calculation of concentration and influence tensors. FE simulations are performed using the Abaqus software \cite{Abaqus2019}. Due to symmetry of the problem, only $1/8$ of the REV is modeled (Fig.~\ref{fig:rev}) which is here discretized using approximately $120,000$ quadratic tetrahedron elements.

\begin{figure}[t]
	\begin{center}
		\includegraphics[width=0.9\linewidth]{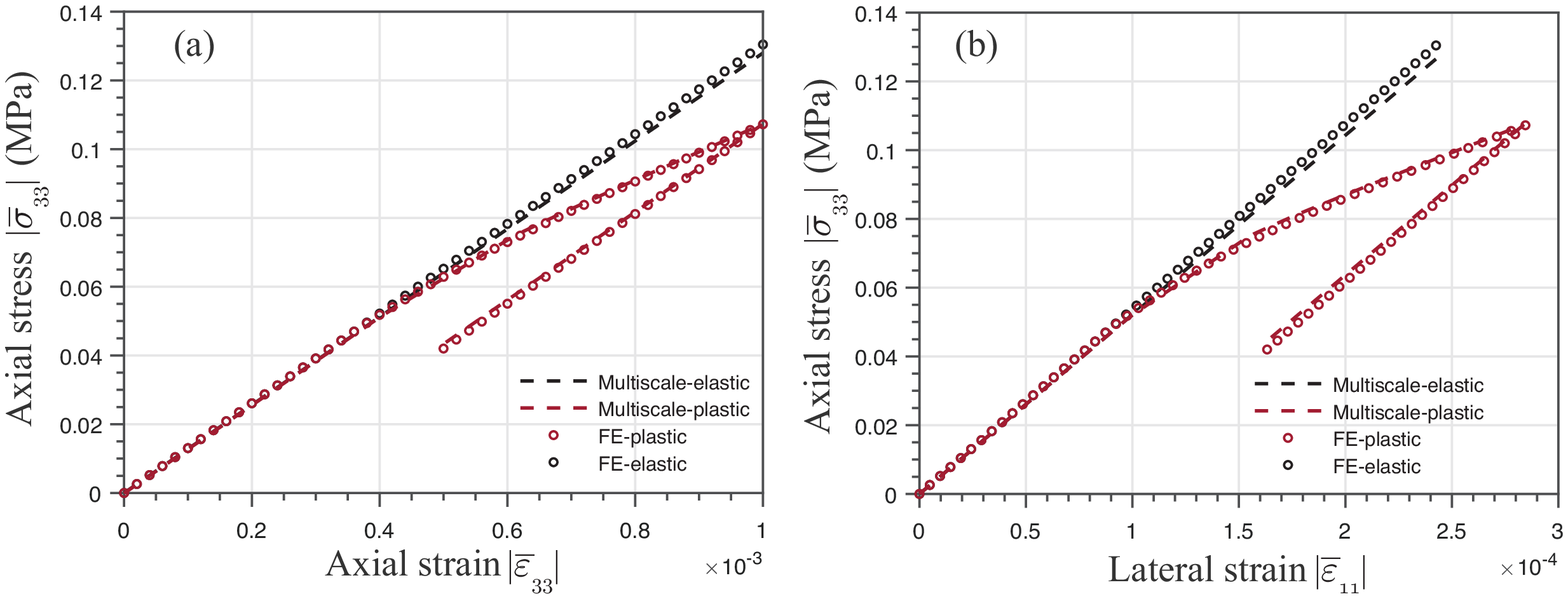}
		\caption{\label{fig:macro-response} Variation of macroscopic axial stress versus macroscopic (a) axial and (b) lateral strains. Comparison between the multiscale model predictions and FE results.}
	\end{center}
\end{figure}

\begin{figure}[h]
	\begin{center}
		\includegraphics[width=0.9\linewidth]{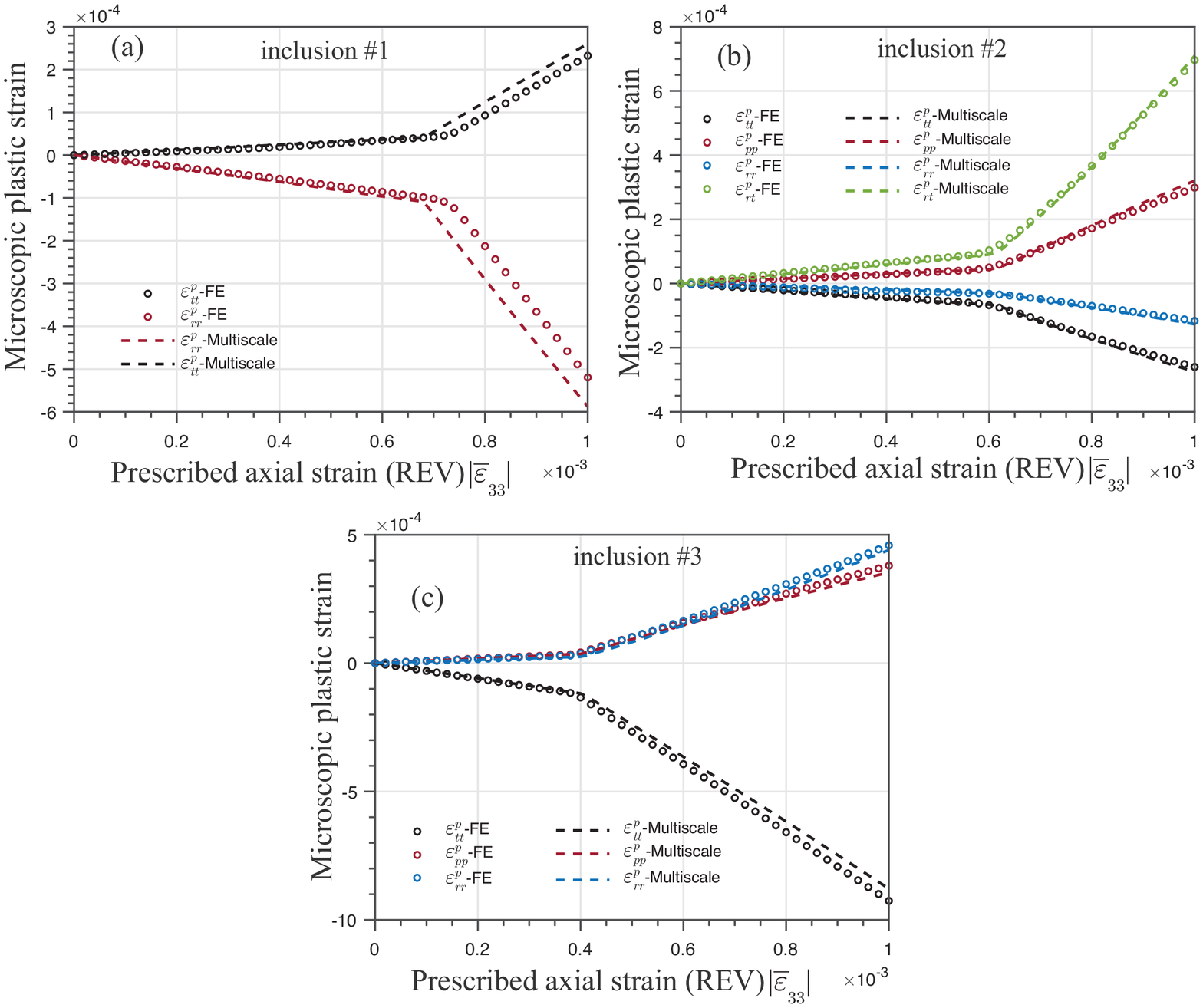}
		\caption{\label{fig:micro-response} Variation of average plastic strain versus prescribed macroscopic axial stress for the three inclusions shown in Fig.~\ref{fig:rev}. The local direction $r$ coincides with the axis of symmetry of the inclusion, while $t$ and $p$ refer to in-plane directions.}
	\end{center}
\end{figure}

Figure~\ref{fig:macro-response} shows the evolution of macroscopic axial stress ($|\overline{\ten{\sigma}}_{33}|$) versus the macroscopic axial ($\overline{\ten{\varepsilon}}_{33}$) and lateral ($\overline{\ten{\varepsilon}}_{11}$) strains. For comparison purposes, the problem is also solved for the case where the inclusions are elastic. It is seen that the multiscale model predicts well the REV response in both the loading and unloading regimes. Next, the components of local average strains for three inclusions (labeled in Fig.~\ref{fig:rev}) are plotted in Fig.~\ref{fig:micro-response} against the prescribed macroscopic axial strain in the loading regime. Slight differences are observed between the model predictions and FE results for inclusions $\#1$ and $\#3$ which are due to the severe interactions between inclusions in the FE model leading to heterogeneous strain/stress fields within inclusions. This is in contrast with the assumption in the multiscale model where the strain (and stress) fields inside inclusions are assumed homogeneous. Nevertheless, it appears that when these discrepancies are volume averaged over the whole REV, they cancel out each other, leading to a good match between the overall responses in the multiscale model and FE (Fig.~\ref{fig:macro-response}).


\begin{figure}[t]
	\begin{center}
		\includegraphics[width=0.475\linewidth]{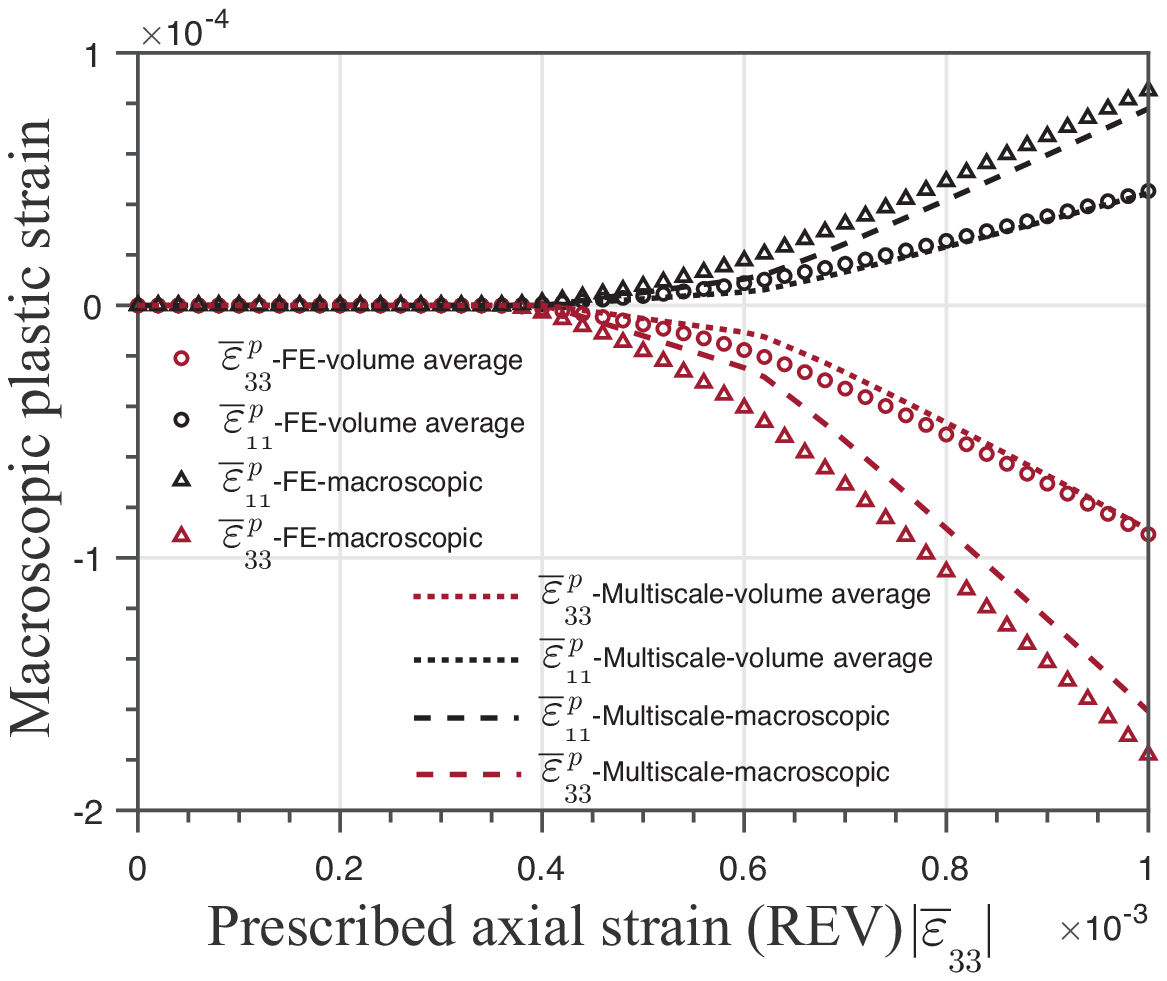}
		\caption{\label{fig:macro-plastic} Volume average of microscopic plastic strain versus the macroscopic ``plastic" strain.}
	\end{center}
\end{figure}

Finally, returning to Eq.~\eqref{eq:avg-pl-strain}, $\overline{\ten{\varepsilon}}^{\text{p}}$ is the macroscopic residual strain that is not recovered upon the removal of the external load on the REV and is classically referred to as ``plastic" strain. It consists of a plastic part which is the volume of average of microscopic plastic strains, and a ``frozen" elastic strain which does not recover in the unloading regime. The macroscopic ``plastic" strain together with its plastic part are plotted in Fig.~\ref{fig:macro-plastic} for both the multiscale and FE models. The model correctly predicts the macroscopic plastic strain demonstrating its capability in modeling the macroscopic unloading regime (as also seen in Fig.~\ref{fig:macro-response}).

\section{Concluding Remarks}
The study investigates the robustness of an eigen-strain based multiscale model formulated within the extended TFA framework for upscaling the elasto-plastic behavior of a heterogeneous REV. Comparing the model predictions with FE simulations of a sample REV shows that such formulation is indeed capable of capturing the macroscopic response of the elasto-plastic REV as well as evolution of microscopic stress and strain fields inside the REV, in both the loading and unloading regimes, with acceptable accuracy. The current formulation thus emerges as a simple, yet powerful, tool for strength predictions of heterogeneous geomaterials.

\begin{acknowledgement}
	This work was funded by the Natural Sciences and Engineering Research Council of Canada (Grant No. RGPIN-2016-04086 held by R.W. and Grants No. RGPIN-2020-06480 and DGECR-2020-00411 held by M.P.).
	
\end{acknowledgement}

\bibliographystyle{spmpsci}
\bibliography{Eghbalian_et_al-Multiscale_Modeling_of_ElastoPlasticity}

\begin{thebibliography}{1}
\providecommand{\url}[1]{{#1}}
\providecommand{\urlprefix}{URL }
\expandafter\ifx\csname urlstyle\endcsname\relax
  \providecommand{\doi}[1]{DOI~\discretionary{}{}{}#1}\else
  \providecommand{\doi}{DOI~\discretionary{}{}{}\begingroup
  \urlstyle{rm}\Url}\fi

\bibitem{Abaqus2019}
{Abaqus 2019}.
\newblock Dassault Systemes Simulia Corporation, Johnston, Rhode Island (2019)

\bibitem{Dvorak1992b}
Dvorak, G.J.: {Transformation field analysis of inelastic composite materials}.
\newblock Proceedings of the Royal Society of London A \textbf{437}(1900),
  311--327 (1992)

\bibitem{Dvorak1992a}
Dvorak, G.J., Benveniste, Y.: {On transformation strains and uniform fields in
  multiphase elastic media}.
\newblock Proceedings of the Royal Society of London A \textbf{437}(1900),
  291--310 (1992)

\bibitem{Eghbalian2019}
Eghbalian, M.: {Hydro-mechanical coupling and failure behavior of argillaceous
  sedimentary rocks: A multi-scale approach}.
\newblock Ph.D. thesis, University of Calgary (2019)

\bibitem{Morin2017}
Morin, C., Vass, V., Hellmich, C.: {Micromechanics of elastoplastic porous
  polycrystals: Theory, algorithm, and application to osteonal bone}.
\newblock International Journal of Plasticity \textbf{91}, 238--267 (2017)

\bibitem{Pichler2010}
Pichler, B., Hellmich, C.: {Estimation of influence tensors for eigenstressed
  multiphase elastic media with nonaligned inclusion phases of arbitrary
  ellipsoidal shape}.
\newblock Journal of Engineering Mechanics \textbf{136}(8), 1043--1053 (2010)

\end{thebibliography}

\end{document}